\documentclass[12pt]{article}

\usepackage[left=2cm,top=3.5cm,right=2cm,bottom=3cm,nohead]{geometry}
\usepackage{amsmath,amsfonts,amsthm,amssymb,times, bm}
\usepackage{epsfig}
\usepackage{color,verbatim,epic, eepic}
\input amssym.def
\input amssym.tex
\usepackage{yfonts}[1998/10/03]

\newfont{\go}{ygoth.tfm scaled 1200}  

\newtheorem{theorem}{Theorem}[section]

\newtheorem{conjecture}[theorem]{Conjecture}

\newcommand{\At}{\widetilde{A}}
\newcommand{\Bt}{\widetilde{B}}
\newcommand{\Ct}{\widetilde{C}}

\newcommand{\dif}[2]{\frac{d #1}{d #2}}

\numberwithin{equation}{section}
\begin{document}

\title{Ricci Flow of Biaxial Bianchi IX Metrics}
\author{Gustav Holzegel$^a$\footnote{G.Holzegel@damtp.cam.ac.uk} \ ,
  Thomas Schmelzer$^{b}$\footnote{Thomas.Schmelzer@balliol.ox.ac.uk} \
  \ \ and \ Claude
  Warnick$^{a}$\footnote{C.M.Warnick@damtp.cam.ac.uk} \\
  \llap{$^a$}{\it \small University of Cambridge, DAMTP, Wilberforce
    Road, CB1 0WA, Cambridge, U.K.} \\  \llap{$^b$}{\it \small University of
  Oxford, Computing Laboratory, Wolfson Building, Parks Road, OX1 3QD,
  Oxford, U.K. }} 
\date{12th June, 2007}

\vspace{1.5cm}

\maketitle

\vspace{-9cm}
\begin{flushright}
\small
DAMTP-2007-53\\
\normalsize
\end{flushright}
\vspace{7cm}

\abstract{We use the Ricci flow with surgery to study 
four-dimensional $SU(2) \times U(1)$-symmetric 
metrics on a manifold with fixed boundary given by a squashed 3-sphere. 
Depending on the initial metric we show that 
the flow converges to either the Taub-Bolt or the Taub-NUT metric,
 the latter case potentially requiring surgery at some 
point in the evolution. The Ricci flow allows us to explore the
Euclidean action landscape within this symmetry class. This work
extends the recent work of Headrick and Wiseman \cite{Wiseman} to more
interesting topologies.}

\section{Introduction}

Ricci flow is a geometric flow within the space of Riemannian
metrics on a given $n$-dimensional manifold $\mathcal{M}$ defined by:
\begin{equation} \label{Ricci} 
\begin{array}{ll}
\dif{}{t} g\left(t,x\right) &= -2 Ric_{g(t,x)} \\
g\left(0,x\right) &= g_0\left(x\right)
\end{array}
\end{equation}
where $g(t)$ is the metric at flow time $t$ and $Ric_{g(t)}$ is its
associated Ricci tensor. On a closed manifold, the PDE 
(\ref{Ricci}) admits short time existence \cite{Hamilton}, 
but will generally exhibit finite time blow-up. Work of Hamilton 
and Perelman has provided a means of continuing the flow beyond 
such a singularity, namely Ricci flow with 
surgery. There has recently been much
interest in the Ricci flow with surgery, especially since the
publication by Perelman of a proof of the Poincar\'{e} conjecture 
based on this method \cite{Perelman1, Perelman2}.

For many practical purposes it is more
convenient to consider a slightly modified Ricci flow, defined by:
\begin{equation} \label{Riccimod} 
\begin{array}{ll}
\dif{}{t} g\left(t,x\right) &= -2 Ric_{g(t,x)} +2 \left(\nabla \xi(t)
\right)_S \\
g\left(0,x\right) &= g_0\left(x\right)
\end{array}
\end{equation}
Here $\xi(t)$ is an arbitrary vector field on $\mathcal{M}$ 
and $\left( \cdot \right)_S$
indicates that the tensor enclosed should be symmetrized on all its
indices. The flows (\ref{Ricci}) and (\ref{Riccimod}) are equivalent 
if we consider the flow to act 
on the space of metrics \emph{modulo diffeomorphisms} as the second term in
(\ref{Riccimod}) on its own would define a flow through the
equivalence class of a metric under diffeomorphisms. The advantage of
(\ref{Riccimod}) however is that one can pick the gauge
$\xi\left(t\right)$, i.e.\ a diffeomorphism at each point of time, 
such that the equations become strongly parabolic. Short-time existence
then follows from standard results \cite{Knopf,deTurck}. Moreover the
stability of numerical codes is improved in this formulation
of the Ricci flow. 

In this paper we are going to study the Ricci flow on a class of
four-dimensional manifolds $\mathcal{M}$ whose boundary is given by a squashed
3-sphere. The physical motivation 
to study infilling metrics for a manifold with fixed boundary 
metric derives from an attempt to understand the
thermodynamics of general relativity. In the Euclidean approach 
to the problem the metric is analytically continued to the Riemannian sector and -- in analogy with
statistical mechanics -- a partition function associated with 
the canonical ensemble is defined by 
\begin{equation} \label{ensemble}
\mathcal{Z}\left(g\right) = \int_\mathcal{M} d\left[g\right] e^{-S\left[g\right]}
\, .
\end{equation}
Here $S\left[g\right]$ is the Euclidean action (\ref{eucact}) and the integration is
performed over all metrics on $\mathcal{M}$
admitting the same fixed boundary metric but (perhaps topologically)
different interiors. The fixed boundary metric can be understood in a
thermodynamic sense as holding the system at fixed temperature. 
However, the integral (\ref{ensemble}) is
not well-defined in general and in the one-loop semi-classical approximation renormalization schemes have to be employed to
extract the correct values for the entropy and other thermodynamic
quantities from the partition function \cite{Gibbons:1978ac}. In practical evaluations of (\ref{ensemble}) one usually restricts to certain 
symmetry classes and invokes a saddle point 
approximation: the main contributions to the integral 
(\ref{ensemble}) will come from extremal points of the action, which
are found to be Ricci-flat metrics. In this setting it is an 
interesting mathematical problem on its own to classify the number 
of infilling Ricci-flat metrics for a given boundary metric. This
question has so far been answered analytically only in restricted symmetry
classes (like the one studied in this paper). On the other hand, 
being a gradient flow, the Ricci flow provides a useful tool to explore the
action landscape of Riemannian metrics. For metrics less
constrained by symmetry assumptions, the Ricci flow may be thought of
as a relaxation technique for finding new Ricci-flat metrics. 

We direct the reader interested in further details, especially 
on the relation of the Ricci-flow to the thermodynamics of black holes 
and the renormalization group in quantum field theory, 
to \cite{Wiseman}.  In that paper the authors study 
the canonical ensemble for gravity in a box, modelled by 
four-dimensional manifolds with a boundary of topology 
$S^1 \times S^2$.\footnote{The
  $S^1$ direction arises from the periodic time
  coordinate of the Lorentzian metric after analytic continuation.}  Furthermore, they 
restrict to metrics admitting a $U(1) \times SO\left(3\right)$
action by isometry.\footnote{Note that symmetry in the initial data 
is preserved under the Ricci flow.} At high temperatures, 
i.e.\ a small enough size of the $S^1$ direction, 
they find three infilling Ricci-flat geometries, which correspond
to saddle points of the action: two 
Schwarzschild black holes of different horizon areas
(with topology $\overline{B^2} \times S^2$) and hot flat space 
(topology $S^1 \times \overline{B^3}$). Considering perturbations of the small 
Schwarzschild solution it is finally shown numerically how the
Ricci-flow converges to either the large black hole or hot flat space,
with the latter case enforcing a surgery at some point. 

The results of the present paper provide a natural extension of the work of 
\cite{Wiseman} to other topologies. Here the boundary will be given by
a squashed $S^3$ and we will restrict attention to manifolds 
endowed with metrics admitting an $SU(2) \times U(1)$ symmetry, 
so called biaxial Bianchi IX metrics. The Ricci flat metrics, i.e.\ the fixed points of the 
flow (\ref{Ricci}), within this class are the one parameter 
families of Taub-Bolt and Taub-NUT metrics, which differ 
in their topology. Taub-Bolt is a metric on $\mathbb{C}P^2 \setminus \{
\textrm{open ball} \}$, whereas Taub-NUT is defined on $\overline{B^4}$. 
We will discuss some properties of these 
metrics, which play a role in Euclidean quantum gravity
\cite{Gibbons:1978ac}, in section 2. 
We also establish that fixing the squashing parameter of the 
3-sphere on the boundary in a certain range in fact 
singles out three infilling Ricci flat
geometries from these families: Two Bolt- and one NUT- solution. 
We demonstrate, by finding a negative eigenvalue of the
linearized operator, that one of these Bolt solutions is 
unstable under the Ricci-flow. The full non-linear evolution of the
unstable solution is finally studied numerically. We find that 
the unstable Bolt metric will flow to either the second (stable) 
Bolt metric or the NUT metric, depending on how the metric is
perturbed initially. Geometrically the perturbation corresponds to whether 
one slightly increases or decreases the area of the small Bolt minimal surface. The crucial and most interesting 
feature, however, is that the latter flow from the Bolt 
to the NUT solution requires surgery because the NUT and Bolt 
solutions have different topology, as
mentioned above. Hence we will explain our method of 
controlling the curvature blow up and how the surgery is finally 
performed. The behaviour under perturbations just described 
might be expected from an analysis of the gravitational 
Euclidean action for the solutions as illustrated below. 

This work, together with that of Headrick and Wiseman, provide
concrete examples of the two types of surgery required by the Ricci
flow in $4$ dimensions as anticipated by Hamilton in \cite{Hamilton2},
which locally look like
\begin{equation}
S^3 \times B^1 \to B^4 \times S^0 \qquad \mathrm{and} \qquad S^2 \times B^2 \to B^3 \times S^1.
\end{equation}
respectively. These surgeries are expected to be irreversible.

To our knowledge the question of existence and uniqueness for the
Ricci flow on manifolds with fixed boundary metric has not 
been answered in any generality. The reason is that 
in the strongly parabolic formulation (\ref{Riccimod}) 
one typically ends up with mixed Dirichlet and Neumann 
conditions at the boundary. We will therefore 
prove uniqueness of the evolution for our Ricci-flow 
equations directly in appendix $A$.

\section{NUTs and Bolts}

\subsection{Regular Biaxial Bianchi IX Metrics}

The biaxial Bianchi-IX metrics are a class of metrics admitting an
$SU(2)_L\times U(1)_R$ symmetry group. They may be written in terms of the
usual left-invariant $su(2)$ one-forms $\sigma_i$ in the form:
\begin{equation}
ds^2 = a(r)^2 dr^2 + b(r)^2 \left(\sigma_1^2 + \sigma_2^2 \right) +
c(r)^2 \sigma_3^2
\end{equation}
The $SU(2)_L$ symmetry arises from the left action of $SU(2)$ under
which the one-forms $\sigma_i$ are by definition invariant. In
addition, since $\sigma_1$ and $\sigma_2$ appear in the metric with
the same coefficient there is a right action by a $U(1)$ subgroup of
$SU(2)$ which acts by isometries. We will be
interested in manifolds with a boundary, so the radial coordinate $r$
will take values in the range $r_0<r\leq r_1$. The boundary at $r_1$ is topologically a $3$-sphere and the induced metric that of a
homogeneously squashed $S^3$. We assume that the functions $a, b, c$
are smooth and non zero on the interval and further that $\{r=r_0\}$
is a fixed point set of the $U(1)_R$ action. This means that
$c(r_0)=0$ and there will in general be a singularity at $r=r_0$
unless certain conditions are met. Without loss of generality, we
assume that $r_0=0$ and
$a(r)^2 \sim a_0^2 + O(r^2)$. If the metric is to be regular at $r=0$
there are two possibilities \cite{Gibbons:1979xm}:

\textbf{Case 1} \ \  $b(0) \neq 0$. In this case regularity requires
that 
\begin{equation}
c(r)^2 \sim \frac{1}{4} a_0^2 r^2 + O(r^4)
\end{equation}
 as $r \to 0$. A coordinate
transformation then shows that $r=r_0$ is a coordinate singularity,
similar to the origin of
$\mathbb{E}^2$ in polar coordinates. The point set $r=0$ is then a
metric $2$-sphere known as a \emph{Bolt}. Geometrically it is a
minimal surface rather than a boundary.

\textbf{Case 2} \ \  $b(0) = 0$. In this case regularity requires that
\begin{equation}
b(r)^2 \sim c(r)^2 \sim \frac{1}{4} a_0^2 r^2 + O(r^4)
\end{equation}
as $r \to r_0$. A
coordinate transformation shows that $r=0$ is a coordinate
singularity, similar to the origin of
$\mathbb{E}^4$ in spherical polar coordinates. The point set $r=0$
is then a point known as a \emph{Nut}.

\subsection{Ricci Flat Metrics}

Clearly from (\ref{Ricci}) the fixed points of the Ricci flow are
metrics satisfying $Ric_g=0$. Such metrics are sometimes known as
gravitational instantons and occur as extremal points of the Euclidean
Einstein-Hilbert action:
\begin{equation}
\label{eucact}
S[g]  = -\frac{1}{16\pi G} \int_{\mathcal{M}} d V
\sqrt{g} R - \frac{1}{8 \pi G}
\int_{\partial \mathcal{M}} d S \sqrt{h}K.
\end{equation}
With a carefully chosen metric on the space of Riemannian metrics, 
the Ricci flow may be thought of as a gradient flow of this action, as
discussed by Headrick and Wiseman \cite{Wiseman}.

\subsubsection{Taub-Bolt}

In order to find Ricci flat infilling metrics for a squashed $S^3$ boundary, we
consider Ricci flat biaxial Bianchi IX metrics whose radial slices are
squashed $S^3$. There are two families of such metrics corresponding to the two possible
regular structures at the fixed points of the $U(1)$ action. The
Taub-Bolt family of metrics is given by:
\begin{equation}
ds^2 = \frac{r^2-n^2}{r^2-\frac{5}{2} n r + n^2}dr^2 +(r^2-n^2) \left(\sigma_1^2+\sigma_2^2\right) +4 n^2 \left(
\frac{r^2-\frac{5}{2} n r + n^2}{r^2-n^2}\right) \sigma_3^2
\end{equation}
This metric is Ricci flat and regular for $r$ in the range $2n<r<\infty$ and there is a
bolt at $r=2n$. In order that this metric may be considered as
infilling a squashed $S^3$ we introduce a boundary by restricting $r$
to $2n <r\leq R$. It will be convenient to change the radial
coordinate and introduce new parameters
$\alpha=\sqrt{2+\frac{3n}{R^2}}, \beta = \sqrt{3}n$ instead of $n,
R$ so that the metric instead takes the form:
\begin{equation}
ds^2 = 8 \beta^2 \alpha^2 \frac{(\alpha^4-\rho^4)}{(\alpha^2-2 \rho^2)^4}
d\rho^2 + \beta^2 \frac{(\alpha^4-\rho^4)}{(\alpha^2-2 \rho^2)^2} \left(
\sigma_1^2+\sigma_2^2\right ) + 2 \beta^2\alpha^2
\frac{\rho^2}{\alpha^4-\rho^4} \sigma_3^2 \label{boltmet}
\end{equation}
Now the radial coordinate is in the range $0<\rho\leq 1$. For
$\alpha^2>2$ this is a regular metric on a manifold with boundary which is topologically
$\mathbb{CP}^2$ with an open $4$-ball removed. In the limit $\alpha^2
\to 2$ the boundary receeds to infinity and we have the asymptotically
locally flat Taub-Bolt metric on $\mathbb{CP}^2 - \{\mathrm{pt.}\}$.

We seek Ricci flat metrics which have a given metric on the
boundary $S^3$, taken to be:
\begin{equation}
h = \mu^2\left (\sigma_1^2 + \sigma_2^2 \right) + \nu^2 \sigma_3^2. \label{bdrymet}
\end{equation}
$h$ can only be the induced metric on the boundary of a Taub-Bolt
solution for certain values of $\tau^2 = \nu^2/\mu^2$, which we refer to as the squashing parameter. If
$\tau^2 > 1-\frac{9}{8}3^{\frac{1}{3}}+\frac{3}{8}3^{\frac{2}{3}}=\tau_c^2$ then
one cannot find a Taub-Bolt solution whose boundary has metric
(\ref{bdrymet}). For $\tau^2<\tau_c^2$ there are two choices of $(\alpha,
\beta)$ such that the boundary metric is $h$. These can be
distinguished via the parameter $\beta$ which parameterises the size of the bolt. The two
solutions are thus distinguished by having a `small bolt' or `large
bolt'. In the critical case where $\tau=\tau_c$, the small and large
bolt solutions coincide and we have only one infilling Taub-Bolt
geometry.

\subsubsection{Taub-Nut}

We now turn to the other family of Ricci flat biaxial Bianchi IX
metrics which may be considered as infilling geometries for the 
boundary $S^3$ with metric (\ref{bdrymet}). These contain a
nut and are known as the Euclidean self-dual Taub-NUT solutions. 
In their usual form they have metric:
\begin{equation}
ds^2 = \frac{r+n}{r-n} dr^2+ 4 n^2 \frac{r-n}{r+n} \sigma_3^2 + \left(
r^2-n^2 \right) \left(\sigma_1^2+\sigma_2^2 \right).
\end{equation}
This metric is regular and Ricci flat for $r$ in the range
$n<r<\infty$. At $r=n$ the metric has a regular nut and we introduce a
finite boundary by restricting $r$ to $n<r\leq R$. As in the
previous case, it will be convenient to change to a
different radial coordinate and new parameters $\gamma=\frac{1}{R} \sqrt{2n}, \delta=\sqrt{R}$ so that the metric takes the form:
\begin{equation}
ds^2 = 4 \delta^2(\rho^2+\gamma^2) d\rho^2 + \delta^2 \rho^2 \left(
(\rho^2 + \gamma^2)(\sigma_1^2+\sigma_2^2)+
\frac{\gamma^4}{\rho^2+\gamma^2} \sigma_3^2 \right) \, , \label{nutg}
\end{equation}
where the radial coordinate is in the range $0<\rho \leq 1$. One can
see by inspection that this metric satisfies the conditions for
$\rho=0$ to be a regular nut. This gives a regular metric on
$\overline{B^4}$, the closed $4$-ball. There is exactly one Taub-Nut
metric which infills the squashed $S^3$ (\ref{bdrymet}) for $\tau$ in
the range $0<\tau<1$.

\subsubsection{Action}

\begin{figure}[!t]
\centering \framebox {\includegraphics[scale=0.7]{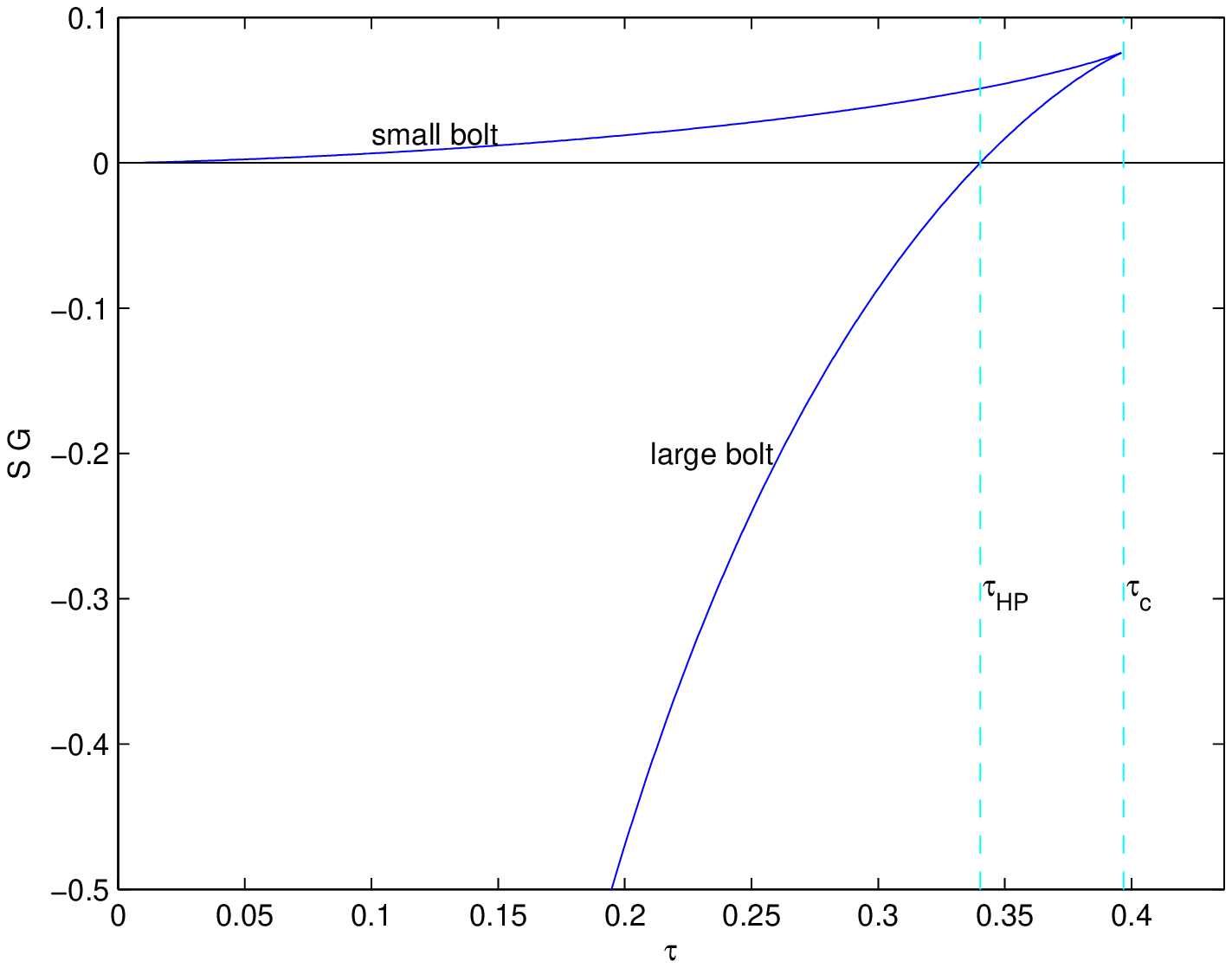}}
\caption{ \label{actplot} A plot of
  $S_{\mathrm{Bolt}}-S_{\mathrm{Nut}}$ as a function of boundary squashing}
\end{figure}

As we know that the Ricci flow is a gradient flow of the Euclidean
action (\ref{eucact}), it is convenient to plot the action of the
Ricci flat solutions. We fix the volume of the boundary metric and
plot the difference $S_{\mathrm{Bolt}}-S_{\mathrm{Nut}}$ against
$\tau$ for both branches of the infilling Taub-Bolt solutions. This
plot is included as Figure \ref{actplot}. We note that the solution
with the greatest action is always the small bolt solution. The
solution with the least action is either the large bolt or the nut,
according as $\tau < \tau_{HP}$ or $\tau > \tau_{HP}$. At
$\tau=\tau_{HP}$ there is a Hawking-Page type transition as the global
minimizer of the action changes from one solution to the other.

In many ways, the Taub-Bolt geometries are similar to the Euclidean
Schwarzschild geometries considered in \cite{Wiseman}. 
After analytic continuation, the horizon of the Schwarzschild 
black hole becomes a regular bolt provided the Wick
rotated time coordinate has some suitable period. If we seek
geometries which infill an $S^2_R \times S^1_\beta$ boundary, then we
find that this can only be done with a Euclidean Schwarzschild metric
for a particular range of the ratio $\beta / R$. As with Taub-Bolt
there is a critical value for this ratio where there is a bifurcation
and two solutions appear. The Schwarzschild solution with a small bolt
has greatest action and is unstable under the Ricci flow, flowing to
either the large bolt solution or to periodically identified flat space.

Motivated by this analogy to Euclidean Schwarzschild, we
expect the small bolt solution to be unstable under Ricci flow in the
case of biaxial Bianchi IX symmetry. More precisely, perturbing the small bolt 
solution the metric should flow to either the large bolt 
solution or the Taub-Nut solution, the latter flow requiring 
a topology changing surgery. We will show in the remainder of the
paper that the flow indeed exhibits the behaviour described.

\section{Ricci flow simulations}

\subsection{Bolt topology -- the linearized problem}

We will begin our simulations by perturbing the small bolt solution as
we expect this to be unstable against the Ricci flow. For the flow on
a manifold with bolt topology we make the metric ansatz:
\begin{equation} 
ds^2 = e^{2 A(r,t)} dr^2 + e^{2 B(r,t)} (\sigma_1^2+\sigma_2^2)+ r^2 e^{2
  C(r,t)} \sigma_3^2 \label{boltans}
\end{equation}
with $0<r\leq 1$. The Taub-Bolt solutions may be cast into this
form. We also make a choice of $\xi$, the vector field which specifies
the gauge in (\ref{Riccimod}). We take the usual $(\theta, \phi,
\psi)$ coordinates for $SU(2)$ and set
\begin{equation}
\xi = \left[-(g_{\mu\nu} \star d \star d x^\nu)+(g_{\mu\nu} \star d
  \star d x^\nu)_0\right] dx^\mu = -(2 B'+C'-A')  dr, \label{boltgauge}
\end{equation}
where the subscript $0$ indicates that the term should be evaluated on
some fixed reference metric, which we take to be (\ref{boltmet}). One
can check that $\xi$ vanishes identically in these coordinates. These are the coordinates
singled out by the gauge choice, which breaks the diffeomorphism
invariance of the Ricci flow equations. This choice of gauge, similar
to that chosen for the `deTurck trick' \cite{deTurck} 
makes the Ricci flow equations
strongly parabolic. We wish to evolve the functions on the interval
$0<r\leq 1$, however $r=0$ is not a boundary for the manifold, but a
point at which the coordinates break down. This is reflected in the
equations we derive from (\ref{boltans}, \ref{boltgauge}) as some
coefficients of the PDE are not bounded as $r\to 0$, cf.~the system
(\ref{eom1})-(\ref{eom3}) in the appendix. The PDE is expected to
admit a well
posed initial value formulation, provided that we impose that the
functions $A, B, C$ extend smoothly to even functions on the interval
$-1 \leq r \leq 1$ as will be discussed in appendix A, where we indeed
prove the uniqueness of solutions with given initial and boundary data. Doubling 
the interval is therefore a convenient strategy to remove the 
apparent boundary at $r = 0$. The pseudo-spectral collocation 
methods \cite{Fornberg} we have used for the numerical simulations 
benefit to a large extent from this idea. We now need
three boundary conditions at $r=1$.\footnote{those for $r=-1$ follow by
  symmetry} Two of these arise by fixing the metric on the
boundary. For well posedness we require a third, which comes from
requiring the vector field $\xi$ to vanish at the boundary. This
ensures that the boundary remains at $r=1$. The boundary conditions
are then:
\begin{eqnarray} \label{abcond}
B(1, t) &=& B(1,0) \nonumber \\
C(1, t) &=& C(1,0)  \\
\xi(1,t) &=& 0 \nonumber
\end{eqnarray}
\begin{figure}[!t]
\centering \framebox {\includegraphics[scale=0.7]{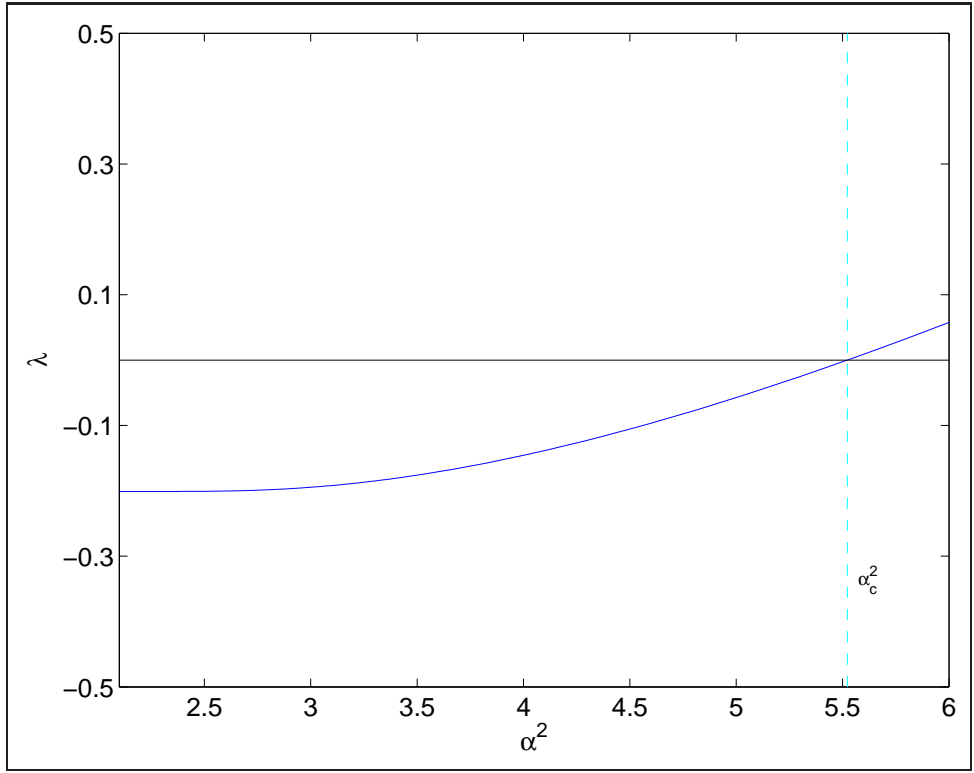}}
\caption{ \label{eigplot} A plot of $\lambda_{min}$ as a function of $\alpha$}
\end{figure}

We expect that the small bolt solution will be unstable under this
flow and that the large bolt solution will be stable. In order to
check this, we linearize the system of equations obtained above about
the known Taub-Bolt solution and seek solutions of the form
\begin{equation}
A(r,t) = A_0(r) + e^{-\lambda t} a(r), \quad B(r,t) = B_0(r) + e^{-\lambda t} b(r), \quad C(r,t) = C_0(r) + e^{-\lambda t} c(r),
\end{equation}
where $a, b, c$ are considered small. It can be shown that $\lambda$ is then an eigenvalue of the
Lichnerowicz Laplacian. If the Lichnerowicz Laplacian has a negative
eigenmode, the perturbations will grow without bound and the solution
will be unstable. We have calculated numerically, using a pseudo-spectral method based on collocation in
Chebyshev points \cite[Chap.~9]{Trefethen} the lowest 
eigenvalue of the Lichnerowicz Laplacian, in the
gauge $\xi$. It is plotted in Figure (\ref{eigplot}) as a function of
$\alpha$, for $\beta^2 = 3$. 
We find, as expected, that the lowest eigenvalue is negative for
$\alpha^2 < \alpha_c^2$ which is the range of parameters corresponding
to the small bolt. The eigenvalue vanishes at the critical point where
the small and large bolt coincide and becomes positive for the large
bolt. With $38$ sampling points, 
the change in sign occurs at the
theoretical value of $\alpha_c^2$ to within $10^{-12}$. Furthermore,
as $\alpha \to 2$, the boundary approaches infinity and we recover the
result of Young \cite{Young:1984dn} that ALE Taub-Bolt has a negative
Lichnerowicz mode with $\lambda \in (-0.200, -0.201)$.\footnote{For
  values of $\alpha^2$ near 2, one must increase the number of
  sampling points to obtain accurate results.} The choice of $\beta$
above is so as to make a direct comparison with Young's work
possible. See also \cite{Warnick} where a cosmological constant is
included. Such negative modes may be interpreted as indicating 
a semi-classical instability of the metric considered as a 
gravitational instanton \cite{Gibbons:1978ac}. They also indicate 
a dynamical instability within general relativity of the
$5$-dimensional ultra-static vacuum metric obtained by adding 
a $-d\tau^2$ term to the metric.

\subsection{Taub-Bolt Topology -- the Initial Flow}

We use the solutions to the linearised problem to seed the flow for
the full non-linear equations. We consider the flow with initial
conditions:
\begin{equation}
A(r) = A_0(r) +\epsilon a_-(r), \quad B(r) = B_0(r) + \epsilon b_-(r), \quad C(r) = C_0(r) + \epsilon c_-(r)
\end{equation}
Where the functions $a_-$ etc. are the negative mode found for the
linearised problem, normalised so that $b_-(0)=1$. In this case a
perturbation with $\epsilon>0$ increases the area of the bolt, while
one with $\epsilon <0$ decreases it.

\begin{figure}[!t]
\begin{minipage}[t]{0.46\linewidth} 
\centering \framebox {\includegraphics[width=3in]{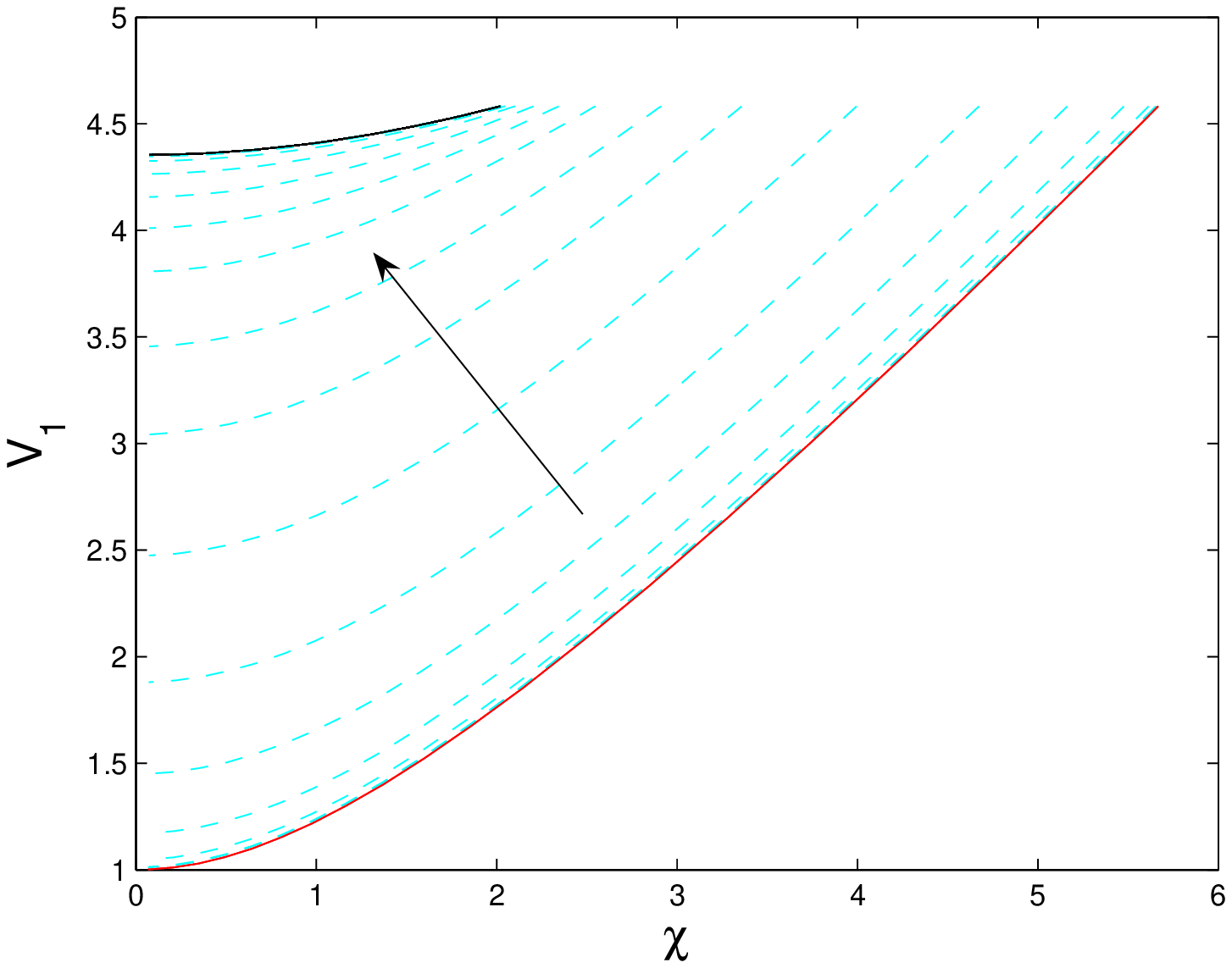}}
\end{minipage}
\hfill 
\begin{minipage}[t]{0.46\linewidth}
\centering \framebox {\includegraphics[width=3in]{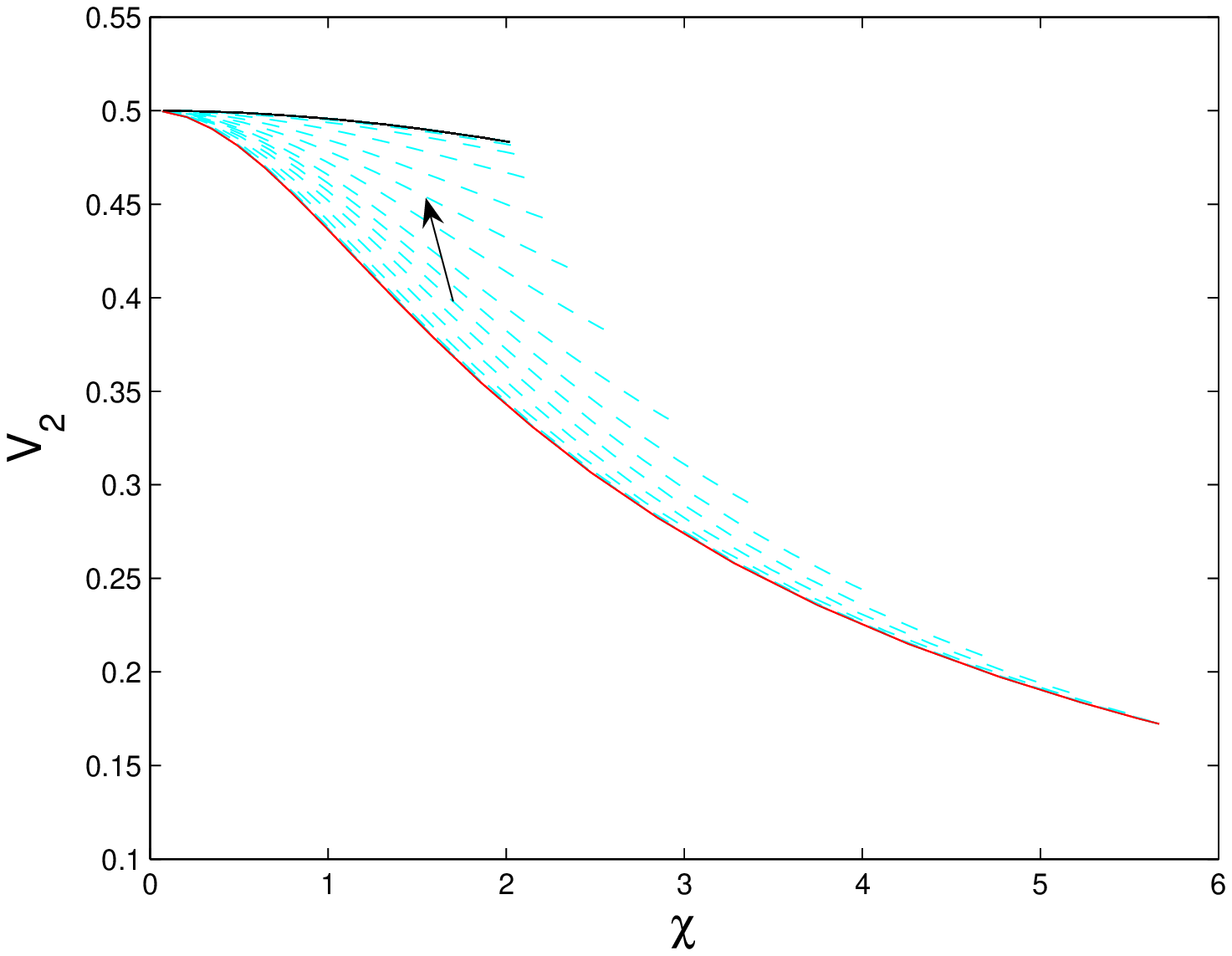}}
\end{minipage}
\caption{Plots of $V_1(\chi)$ and $V_2(\chi)$ at selected times during the flow for $\beta = 1, \alpha^2 = 2.5$. The
  flow proceeds in the direction of the arrow. The two infilling bolt
  geometries are shown as solid lines.}
\end{figure}

We first consider perturbing the small bolt solution with $\epsilon
>0$ which produces a small increase in the area of the bolt. The PDEs
were again discretized in space using pseudo-spectral collocation in Chebyshev 
points. The time-stepping was performed with MATLAB's \texttt{ode15s} 
numerical integrator. We find that for a
range of values of $\alpha^2$ and $\epsilon>0$ the
metric at time $t$ approaches the large bolt solution as $t$ becomes
large. In order to plot the results, we first cast the metric in the
gauge invariant form:
\begin{equation}
ds^2 = d\chi^2 + V_1(\chi)^2 \left(\sigma_1^2+\sigma_2^2 \right) +
\chi^2 V_2(\chi)^2 \sigma_3^2
\end{equation}
and then plot the functions $V_1$ and $V_2$. We see that the flow
starting at the small bolt converges to the large bolt. Taking $102$
spatial sampling points and $\beta=1, \alpha^2=2.5$, we find that for $t > 40$ the
functions have uniformly approached the large bolt solution to within
$10^{-9}$, with better accuracy for more sampling
points.\footnote{However, there is a limit due to the fact that for
  very large $N$ the matrices typically become ill-conditioned.} 
For larger values of $\alpha^2$ we also get better 
accuracy, indeed for $\alpha^2=4$ we can achieve $10^{-11}$ 
accuracy with only $42$ sampling points.

We now turn to the situation where $\epsilon <0$. In this case, the
numerical simulations appear to exhibit a finite time blow-up which
occurs because the bolt becomes smaller and smaller and approaches
zero size in finite time. This behaviour where an $S^2$ collapses to a
point in finite time is characteristic of the Ricci flow. In order to
continue the flow, we must perform a surgery, turning the bolt at the
origin to a nut. To decide when such a surgery should be
performed, we monitor the value of the coordinate invariant scalar
$K = Riem^2|_{r=0}$. We stop the flow when $t = t_c$, defined by $K(t_c) > 10^8$.

\subsection{Taub-NUT topology -- Surgery and Beyond}

The idea behind surgery for the Ricci flow is that given a solution to
the Ricci flow equations exhibiting finite time blow-up at a time $T$,
we should take the Riemannian manifold at a time $T-\delta$ and remove a small region of
size $\epsilon$ from the manifold and glue in a new piece which has a
different topology and smoothly extends the metric on the rest of the
manifold. We then re-start the Ricci flow. If this is done
appropriately, the resulting flow should not depend on the details of
the gluing in the limit $\delta \to 0$. 

\begin{figure}[!t]
\begin{minipage}[t]{0.46\linewidth} 
\centering \framebox {\includegraphics[width=3in]{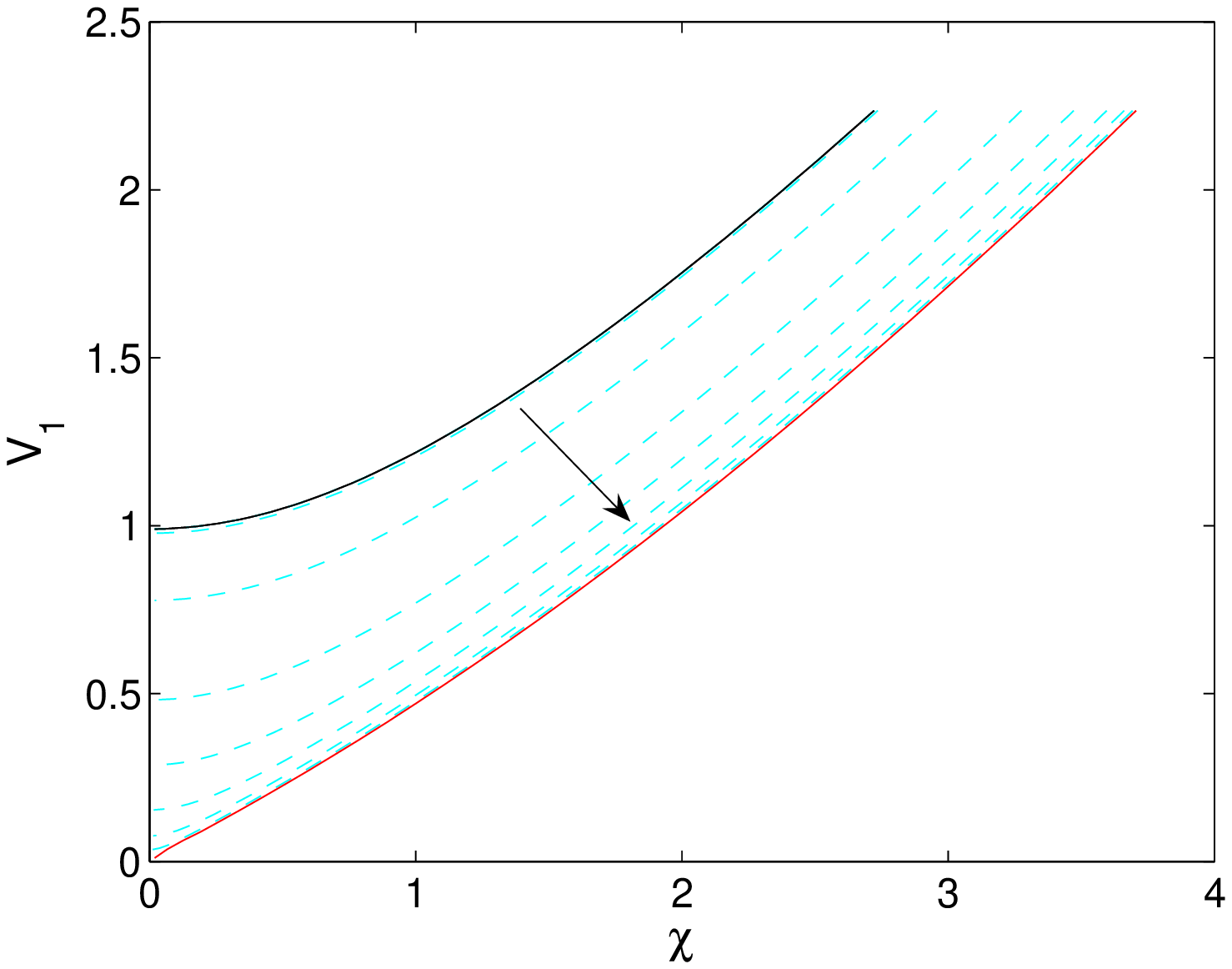}}
\end{minipage}
\hfill 
\begin{minipage}[t]{0.46\linewidth}
\centering \framebox {\includegraphics[width=3in]{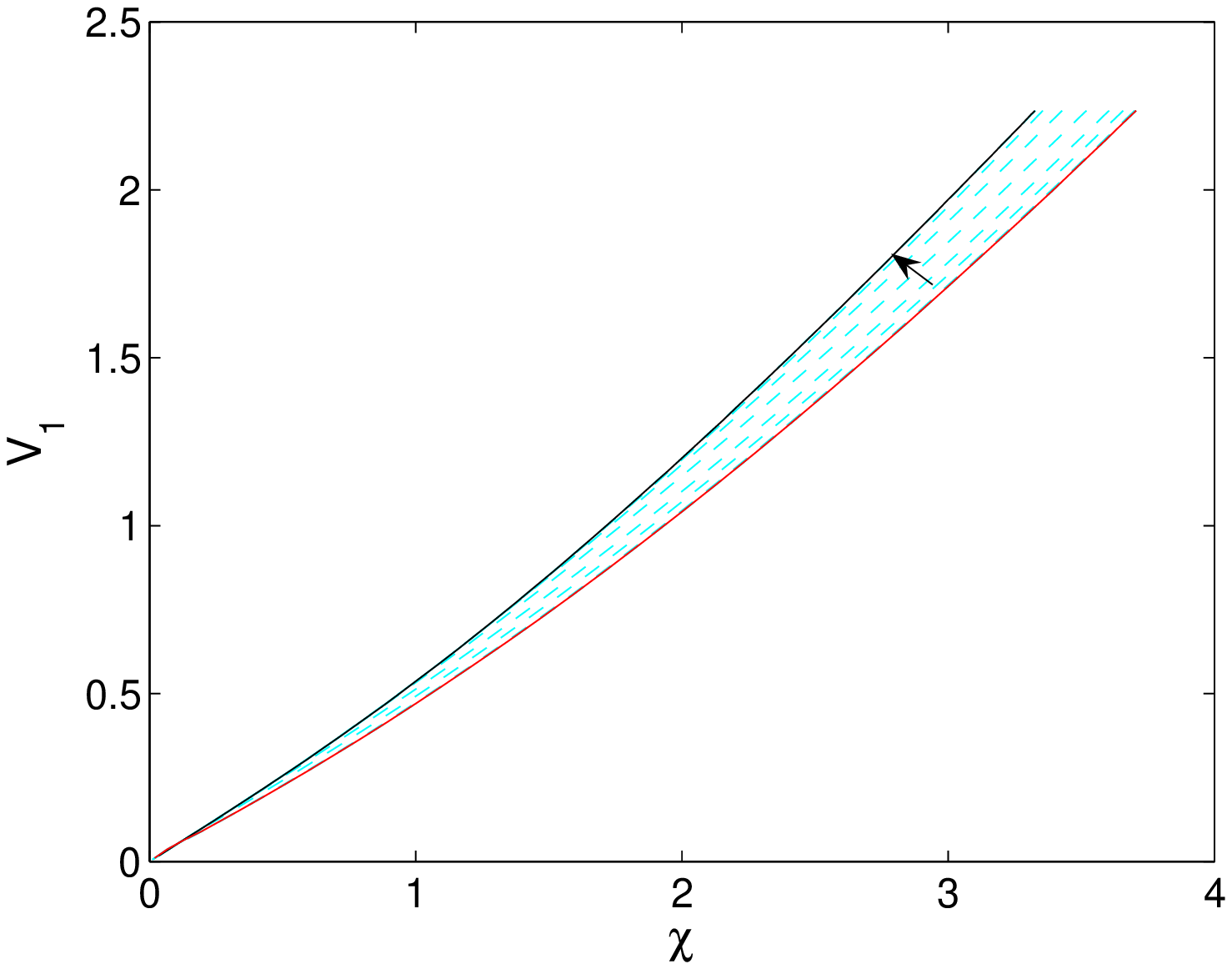}}
\end{minipage}
\caption{Plots of $V_1(\chi)$ at various times, both before surgery
  (on the left) and after surgery (on the right). The solid red line
  shows the function after surgery on both plots. The solid black
  lines are the functions for Taub-Bolt on the right and Taub-NUT on
  the left. \label{V1nut}}
\end{figure}

In our case, the minimal
$S^2$ at $r=0$ appears to be shrinking to zero size in a finite
time. In order to continue the flow beyond this singularity, we remove
the region $r< \epsilon$ from the manifold and glue in a
$4$-ball. This means that the new metric will take the form:
\begin{equation}
ds^2 = e^{2 \widetilde{A}}\left(dr^2 +\frac{r^2}{4} e^{-2 r^2 \widetilde{C}} \left( e^{2 r^2
  \widetilde{B}}(\sigma_1^2 + \sigma_2^2)+\sigma_3^2 \right) \right )
\end{equation}
We require that this agrees with the metric at the point we stopped
the flow,  $g(t_c)$,  for $r>\epsilon$
and is everywhere smooth. This leaves considerable ambiguity in
prescribing the metric functions for $r<\epsilon$, however numerical
experimentation showed that the long term properties of the flow are
insensitive to the precise gluing technique. 

We now wish to continue the Ricci flow with the new initial conditions
given by the metric functions after surgery. We will again introduce a
gauge field $\widetilde{\xi}$ to make the Ricci flow equations strongly
parabolic. We make a similar choice of gauge, given by:
\begin{equation} \label{xinut}
\widetilde{\xi} =\left[-(g_{\mu\nu} \star d \star d x^\nu)+(g_{\mu\nu} \star d
  \star d x^\nu)_0\right] dx^\mu= -\left[ 2 \widetilde{A}'+2 (r^2 \Bt
  )'-3(r^2 \Ct )' \right ]  dr. 
\end{equation}
This time we use the Taub-NUT metric in coordinates (\ref{nutg}) as a
reference metric. The equations which arise are once again symmetric
under $r \to -r$ and so we can set the boundary conditions at the
origin by requiring that the functions $\widetilde{A}, \widetilde{B},
\widetilde{C}$ extend smoothly to an even function on $[-1,1]$. The
symmetric boundary conditions are then given by:
\begin{eqnarray} \label{abcnutond}
(\At-\Ct+\Bt)(1, t) &=& (\At-\Ct+\Bt)(1,0) \nonumber \\
(\At-\Ct)(1, t) &=& (\At-\Ct)(1,0) \nonumber \\
\widetilde{\xi}(1,t) &=& 0
\end{eqnarray}
which indicate that the boundary metric is fixed and the
diffeomorphism $\widetilde{\xi}$ fixes $r=1$. This final condition is
not satisfied by the functions obtained via surgery, however a near
identity change of the $r$ coordinate can make the initial conditions
compatible with the boundary conditions.

We can show (see Appendix) that the system of PDEs obtained from
(\ref{Riccimod}) with the gauge choice (\ref{xinut}) and boundary
conditions (\ref{abcnutond}) has the uniqueness property. We believe
that short term existence also holds, making the problem well
posed. One might think that the parity condition could be replaced by
$\At'(0,t)=0$ etc.\ however it appears that with this boundary
condition the system is ill posed and short term solutions do not
exist for generic initial data.

\begin{figure}[!t]
\begin{minipage}[t]{0.46\linewidth} 
\centering \framebox {\includegraphics[width=3in]{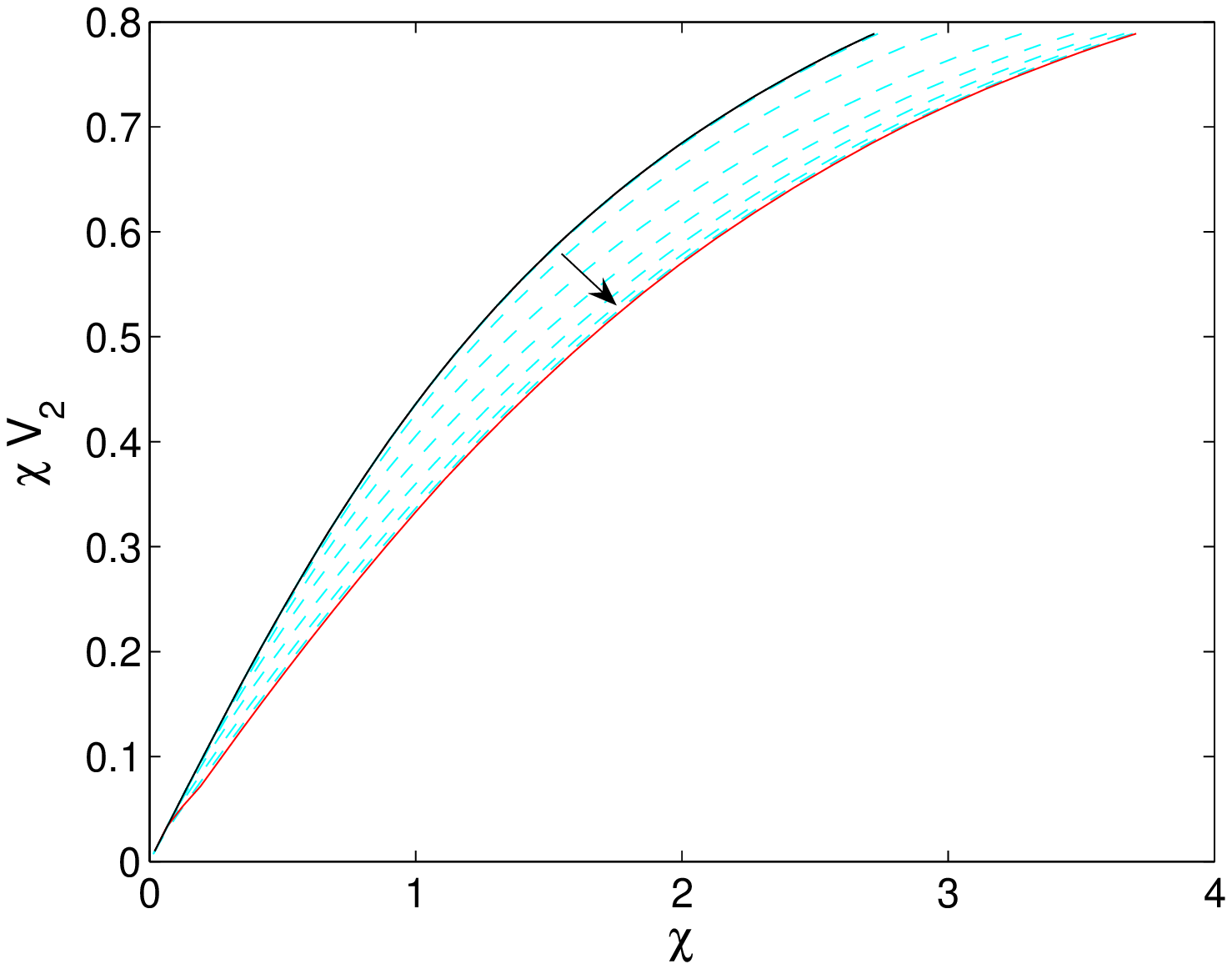}}
\end{minipage}
\hfill 
\begin{minipage}[t]{0.46\linewidth}
\centering \framebox {\includegraphics[width=3in]{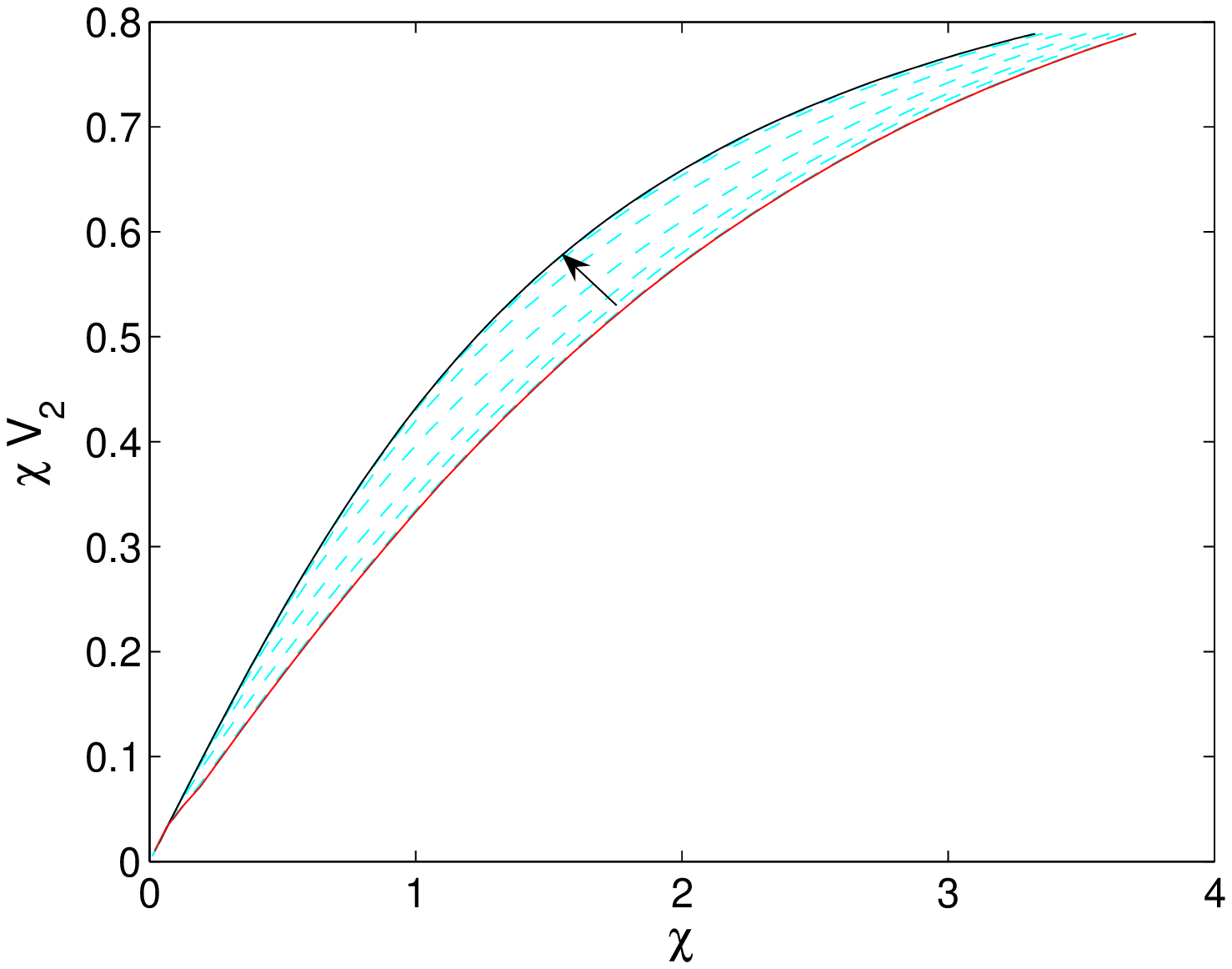}}
\end{minipage}
\caption{Plots of $\chi V_2(\chi)$ at various times, both before surgery
  (on the left) and after surgery (on the right). The solid red line
  shows the function after surgery on both plots. The solid black
  lines are the functions for Taub-Bolt on the right and Taub-NUT on
  the left.\label{V2nut} }
\end{figure}

The Ricci flow equations were solved numerically both before and
after surgery using the same technique as for the flow not requiring
surgery, starting from a Taub-Bolt metric with $\beta = 1$, $\alpha^2 = 3.5$. The results are shown in figures \ref{V1nut} and
\ref{V2nut}. For these plots it is more convenient to plot $V_1$ and
$\chi V_2(\chi)$, where the functions $V_i$ are defined as in the
previous section. We see that although the curvature blows up at $r=0$
under the flow within the bolt topology, the gauge invariant functions
are well behaved and the change due to the surgery is slight. We find
that for sufficiently many sampling points the final functions
approach the functions of Taub-NUT uniformly to within
$10^{-6}$. Experimentation with arbitrary initial conditions suggests
that the Taub-NUT metric is the final state of the flow for any metric
within the $\overline{B^4}$ topology with appropriately squashed $S^3$ boundary conditions.

\section{Conclusion and Open Questions}
In this paper we explored the Euclidean action landscape of 
biaxial Bianchi IX metrics using the Ricci flow with surgery. We 
demonstrated the existence of three saddle points of 
the Euclidean action and showed numerically 
how the Ricci-flow takes us from one to the other, namely from the 
unstable Bolt-solution to either the stable Bolt or the
NUT-solution depending on the perturbation. 

It is an ambitious analytical problem to make the observations 
of the present paper rigorous, which would involve applying the 
methods of Hamilton and Perelman to study the long-time behaviour of
the Ricci-flow. We finish the paper with two conjectures, which we
hope to address rigorously in the future. The first expresses our 
belief that we have exhausted the qualitative features of the Ricci-flow 
within this symmetry class by our study:
\begin{conjecture}
For arbitrary biaxial Bianchi-IX symmetric initial-data and fixed
metric on the boundary $S^3$ given by (\ref{bdrymet}) with $0<\tau<1$, 
Ricci-flow plus at most one surgery will converge to a member of 
either the Taub-Bolt or the Taub-NUT family. 
\end{conjecture}
Restricting to the subcase of NUT-topology, one might -- encouraged by
the final remark of the previous section -- hope to prove
the stronger
\begin{conjecture}
For any biaxial Bianchi-IX symmetric initial-data with Taub-NUT
topology $\left(\overline{B^4}\right)$, the Ricci-flow admits
long-time existence and converges to a member of the Taub-NUT family.
\end{conjecture}
It should be interesting to see how the different geometric features 
of the Bolt and the Nut make their way into the estimates necessary to
establish the long-time behaviour of the flow. 

Let us finally mention possible extensions of the present
work. With the tools at hand it should be possible to include a
cosmological constant into the formulae and impose AdS boundary
conditions. We expect similar results to hold with Taub-Bolt and
Taub-NUT being replaced by their AdS generalizations. 

Moreover, there is a topological generalization if one
considers the more general Lens spaces instead of the squashed 
$S^3$ to function as the fixed boundary metric. For the
case of $\mathbb{R}P^3$ we expect the Eguchi-Hanson metric to be
a stable infilling solution. 

\section{Acknowledgements}
We would like to thank Mihalis Dafermos, Gary Gibbons, Christoph
Ortner, Julian Sonner and David Tong for useful discussions. 
Moreover, we are grateful to the following funding bodies 
for financial support: Studienstiftung des deutschen 
Volkes and EPSRC (GH), PPARC (CMW) and the Rhodes Trust (TS).

\appendix
\section{Uniqueness}
In this appendix we are going to prove uniqueness of the solution to
our modified Ricci-flow equations. As we noted in the introduction, 
short-time existence and uniqueness has been proven for the Ricci 
flow in general on closed manifolds. On the other hand, for manifolds 
with boundary the only general existence and uniqueness result we know 
of is \cite{Shen} where the Neumann problem
\begin{eqnarray} \label{Ricflow3}
\partial_{t} g_{ij} \left(x,t\right) &=& -2R_{ij}\left(x,t\right) \nonumber \\
g_{ij} \left(x,0\right) &=& \hat{g}_{ij}\left(x\right) \\  \nonumber
K_{ab} &=& 0\textrm{\ \ \  on $\partial\mathcal{M}^{n}$}
\end{eqnarray}
with $K_{{ab}}$ being the extrinsic curvature on the boundary, is
studied. It turns out that the boundary has to satisfy certain
geometric conditions in order for (\ref{Ricflow3}) to admit short-time
existence: For a totally geodesic boundary existence and uniqueness
can be shown \cite{Shen} by an application of the general 
results on parabolic boundary value problems described in \cite{Lady}.

The Dirichlet problem, the problem of fixing the metric itself on the
boundary, has not yet been considered in generality. The reason seems
to be that the deTurck technique will always introduce
a derivative condition at the boundary and the resulting boundary
conditions for the modified flow are of non-standard type. Note that
our boundary condition (\ref{abcond}) is typical in that sense.
 
This motivates a more detailed study of existence and uniqueness
within the biaxial symmetry class considered in the paper. Here we
will prove uniqueness of the modified Ricci-flow. A more practical
reason which originally initiated the study was the occurrence of 
numerical solutions blowing up near the origin in our evolution. 
It turns out that in addition to the boundary conditions a topological 
condition has to be imposed near the origin corresponding to the
avoidance of a conical singularity. Without this additional condition 
our uniqueness proof would not go through.\footnote{Uniqueness is
  likely to be violated if that condition is not imposed, as
  illustrated by the  following lower dimensional elliptic example. 
  The infilling Ricci-flat geometries for a manifold with fixed metric
  on a $S^1$ boundary are the flat disc but also the cone. This type
  of non-uniqueness is expected to carry over to the parabolic case.} \\
The Ricci-flow equations for the metric ansatz (\ref{boltans})
constitute the following non-linear system
\begin{eqnarray}
\dot{A} &=& e^{-2A} A'' - \frac{1}{r} e^{-2A} A^\prime +
\frac{2}{r}e^{-2A} C^\prime + e^{-2A} \left(C'^2 +2 B'^2 - A'^2
\right) \label{eom1} \\
\dot{B} &=& e^{-2A} \cdot B^{\prime \prime} + \frac{1}{r} e^{-2A} B^\prime
 - e^{-2B} + \frac{1}{2} r^2 e^{-4B+2C} \label{eom2} \\
\dot{C} &=& e^{-2A} \cdot C^{\prime \prime} + \frac{1}{r} e^{-2A}
C^\prime -\frac{1}{2} r^2 e^{-4B+2C} \label{eom3}   
\end{eqnarray}
The variable $r$ ranges from $0$ to $1$. We will prove uniqueness
within the space of smooth \emph{even} functions. In particular, 
the Neumann conditions, $A'(t,0)=0$, $B'(t,0)=0$, $C'(t,0)=0$ hold at
$r=0$. At $r=1$ the Dirichlet conditions, $B(t,1)=B_{0}(1)$,
$C(t,1)=C_0(1)$, and the Neumann condition, $\left(2B^\prime +C^\prime -
A^\prime\right)\left(t,1\right)=0$ is imposed, as was 
outlined in section 2. Moreover, to avoid a conical singularity at the
origin, we must have $\lim_{r \rightarrow 0} \left(A-C\right) = \log{2}$. \\ 

To avoid the coupled boundary condition we make the transformation 
\begin{equation}
E = 2B+C-A
\end{equation}
to find the system
\begin{eqnarray}
e^{4B+2C-2E} \dot{B} &=& B^{\prime \prime} + \frac{1}{r} B^\prime
 - e^{2B+2C-2E} + \frac{1}{2}r^2 e^{4C-2E}  \\
e^{4B+2C-2E} \dot{C} &=& C^{\prime \prime} + \frac{1}{r} C^\prime 
-\frac{1}{2} r^2 e^{4C-2E}  \\
e^{4B+2C-2E} \dot{E} &=& E'' + \frac{1}{r} E^\prime +
\left(-\frac{2}{r} E^\prime+\frac{4}{r} B^\prime\right)  \\
&& -\left(C'^2
+2 B'^2 - \left[2B'+C'-E'\right]^2 \right) + \frac{1}{2}r^2 e^{4C-2E}
- 2 e^{2B+2C-2E} \nonumber
\end{eqnarray}
where now $E'=0$ at the boundary $r=1$. 
Next we assume the existence of two even solutions $\left(B,C,E\right)$ and
$\left(\tilde{B},\tilde{C},\tilde{E}\right)$, both bounded 
with all derivatives up to order $k$, say, in $[0,1] \times [0,T]$ for
fixed $T$. We are going to consider the equations for their difference
\begin{equation}
\left(\mathcal{B}, \mathcal{C}, \mathcal{E}\right) = \left(B-\tilde{B}, C-\tilde{C}, E-\tilde{E}\right)
\end{equation}
The equations are
\begin{eqnarray}
e^{4B+2C-2E} \dot{\mathcal{B}} &=& \mathcal{B}^{\prime \prime} +
\frac{1}{r} \mathcal{B}^\prime +
\dot{\tilde{B}}\left(e^{4\tilde{B}+2\tilde{C}-2\tilde{E}}-e^{4B+2C-2E}\right)
\nonumber \\
&&- \left(e^{2B+2C-2E}-e^{2\tilde{B}+2\tilde{C}-2\tilde{E}}\right) +
\frac{1}{2}r^2 \left(e^{4C-2E}-e^{4\tilde{C}-2\tilde{E}}\right)
\label{eq1} \\
e^{4B+2C-2E} \dot{\mathcal{C}} &=&  \mathcal{C}^{\prime \prime} +
\frac{1}{r} \mathcal{C}^\prime + \dot{\tilde{C}}\left(e^{4\tilde{B}+2\tilde{C}-2\tilde{E}}-e^{4B+2C-2E}\right) \nonumber \\
&& -\frac{1}{2}r^2 \left(e^{4C-2E}-e^{4\tilde{C}-2\tilde{E}}\right)
\label{eq2} \\
e^{4B+2C-2E} \dot{\mathcal{E}} &=& \mathcal{E}'' + \frac{1}{r} \mathcal{E}^\prime +
\left(-\frac{2}{r} \mathcal{E}^\prime+\frac{4}{r}
\mathcal{B}^\prime\right) +
\dot{\tilde{E}}\left(e^{4\tilde{B}+2\tilde{C}-2\tilde{E}}-e^{4B+2C-2E}\right)
\nonumber \\
&& - \Bigg(\left[C+\tilde{C}\right]'\mathcal{C}^\prime
+2\left[B+\tilde{B}\right]'\mathcal{B}^\prime \nonumber \\ 
&& \phantom{XX}- \left[2B+C-E +
  2\tilde{B} + \tilde{C} - \tilde{E}\right]^\prime
\left(2\mathcal{B}^\prime + \mathcal{C}^\prime - \mathcal{E}^\prime\right)\Bigg) \nonumber
\\ && + \frac{1}{2}r^2 \left(e^{4C-2E}-e^{4\tilde{C}-2\tilde{E}}\right)
- 2 \left(e^{2B+2C-2E}-e^{2\tilde{B}+2\tilde{C}-2\tilde{E}}\right)  \label{eq3}
\end{eqnarray}
with the initial data
\begin{eqnarray}
\mathcal{B}(0,r)=\mathcal{C}(0,r)=\mathcal{E}(0,r)=0
\end{eqnarray}
and boundary conditions
\begin{eqnarray}
\mathcal{B}(t,1)=\mathcal{C}(t,1)=0 \textrm{ \ \ \ and \ \ \ } \mathcal{E}^\prime(t,1)=0
\end{eqnarray}
as well as the conical condition
\begin{equation}
\mathcal{E}-2\mathcal{B} \sim r^2 \textrm{\ \ \ at the origin}
\end{equation}
plus regularity conditions at the origin arising from the fact
that the functions are even. \\
We now multiply equation (\ref{eq1}) by
$\left(r \mathcal{B} + r \dot{\mathcal{B}}\right)$, (\ref{eq2}) by $r\mathcal{C}$, (\ref{eq3}) by
$\frac{1}{k}r\left(\mathcal{E}-2\mathcal{B}\right)$, add the terms up and integrate over time and space. Here $k$ is a fixed constant which we are going to determine below ($k=16$ will do for instance). On the left hand side we obtain for the $\mathcal{B}$ equation (\ref{eq1}) when multiplied by $\left(r \mathcal{B} + r \dot{\mathcal{B}}\right)$ and integrated
\begin{eqnarray} \label{lhs1}
\int_{t_1}^{t_2} \int_0^1 r \left[e^{4B+2C-2E} \left(
  \dot{\mathcal{B}}\mathcal{B} + {\dot{\mathcal{B}}}^2\right)\right] dr dt =
  \nonumber \\
  \frac{1}{2} \int_0^1 r e^{4B+2C-2E}
  \mathcal{B}^2 dr \Big|_{t=t_1}^{t=t_2} \nonumber \\ + \int_{t_1}^{t_2} \int_0^1 r
  e^{4B+2C-2E} {\dot{\mathcal{B}}}^2 dr dt \nonumber \\ + \frac{1}{2} \int_{t_1}^{t_2} \int_0^1 r
  e^{4B+2C-2E} \left(-4\dot{B} - 2\dot{C} + 2\dot{E} \right) \mathcal{B}^2 dr dt
\end{eqnarray}
for the $\mathcal{C}$ equation (\ref{eq2}) when multiplied by $\mathcal{C}r$ and integrated we find
\begin{eqnarray} \label{lhs2}
\int_{t_1}^{t_2} \int_0^1 r \left[e^{4B+2C-2E} \left(\mathcal{C}
  \partial_t \mathcal{C}\right)\right] dr dt =
  \nonumber \\
  \frac{1}{2} \int_0^1 r e^{4B+2C-2E}
  \mathcal{C}^2 dr \Big|_{t=t_1}^{t=t_2} \nonumber \\ + \frac{1}{2} \int_{t_1}^{t_2} \int_0^1 r
  e^{4B+2C-2E} \left[ \left(-4\dot{B} - 2\dot{C} + 2\dot{E} \right) \mathcal{C}^2 \right] dr dt
\end{eqnarray}
and finally for the left hand side of the $\mathcal{E}$ equation (\ref{eq3}) after multiplication by $\frac{1}{k}r\left(\mathcal{E}-2\mathcal{B}\right)$ and integration
\begin{eqnarray} \label{lhs3}
\int_{t_1}^{t_2} \int_0^1 \frac{1}{k} r \left[e^{4B+2C-2E} \left( \partial_t \mathcal{E}\right)\left(\mathcal{E}-2\mathcal{B}\right) \right] dr dt =
  \nonumber \\
  \frac{1}{2k} \int_0^1 r e^{4B+2C-2E} \left(\mathcal{E}^2 - 4 \mathcal{E} \mathcal{B} \right) dr \Big|_{t=t_1}^{t=t_2} \nonumber \\ + \int_{t_1}^{t_2} \int_0^1 \frac{1}{2k} r
  e^{4B+2C-2E} \left[ \left(-4\dot{B} - 2\dot{C} + 2\dot{E} \right) \left(\mathcal{E}^2-4\mathcal{E}\mathcal{B}\right) \right] dr dt \nonumber \\
+ \int_{t_1}^{t_2} \int_0^1 \frac{1}{k} r \left[e^{4B+2C-2E} \left( 2\mathcal{E} \dot{\mathcal{B}} \right) \right] dr dt 
\end{eqnarray}
The spacetime integrals of equations (\ref{lhs1}), (\ref{lhs2}), (\ref{lhs3}) not containing a time-derivative of $\mathcal{B}$, $\mathcal{C}$ or $\mathcal{E}$ are controlled by  
\begin{eqnarray}
K \int_{t_1}^{t_2} \int_0^1 r e^{4B+2C-2E}
  \left(\mathcal{B}^2 + \mathcal{C}^2 + \mathcal{E}^2 \right) dr dt
\end{eqnarray}
where $K$ is a large constant. This follows from the fact that we control the time-derivatives of the
functions $A, B$ and $C$ by some constant. Note also that the term in the last line of (\ref{lhs3}) can be bounded by
\begin{equation}
\int_{t_1}^{t_2} \int_0^1 \frac{1}{k} r \left[e^{4B+2C-2E} \left( 2\mathcal{E} \dot{\mathcal{B}} \right) \right] dr dt \leq \int_{t_1}^{t_2} \int_0^1 \frac{1}{k} r \left[e^{4B+2C-2E} \left(\mathcal{E}^2 + \dot{\mathcal{B}}^2 \right) \right] dr dt  
\end{equation}
Let us turn to the terms which are produced on the right hand side when equation (\ref{eq1}) is multiplied by $\left(r \mathcal{B} + r \dot{\mathcal{B}}\right)$ followed by a spacetime integration. For the highest order derivative terms we obtain
\begin{eqnarray}
\int_{t_1}^{t_2} \int_0^1 \left(\mathcal{B}'' + \frac{1}{r} \mathcal{B}' \right) \left(\mathcal{B} + \dot{\mathcal{B}}\right) r dr dt = \int_{t_1}^{t_2} r \left(\mathcal{B}' \mathcal{B} + \mathcal{B}' \dot{\mathcal{B}} \right) \Big|^{r=1}_{r=0} dt \nonumber \\
- \frac{1}{2} \int_0^1 \left(\mathcal{B}'\right)^2 r dr \Big|_{t_1}^{t_2} - \int_{t_1}^{t_2} \int_0^1 r \left(\mathcal{B}^\prime\right)^2 dr dt
\end{eqnarray}
The boundary term vanishes because $\mathcal{B}$ satisfies Dirichlet conditions at $r=1$ and Neumann conditions at $r=0$ (even function). Analogously we estimate the highest order terms appearing on the right hand side of equation (\ref{eq2}) when multiplied by $\mathcal{C}r$:
\begin{eqnarray}
\int_{t_1}^{t_2} \int_0^1 \left(\mathcal{C}'' + \frac{1}{r} \mathcal{C}' \right) \mathcal{C} r dr dt = \int_{t_1}^{t_2} r \mathcal{C}' \mathcal{C}  \Big|^{r=1}_{r=0} dt - \int_{t_1}^{t_2} \int_0^1 r \left(\mathcal{C}^\prime\right)^2 dr dt
\end{eqnarray}
where again the boundary term vanishes. Finally, for the right hand side of equation (\ref{eq3}) being multiplied by $\frac{1}{k}r\left(\mathcal{E}-2\mathcal{B}\right)$  we use
\begin{eqnarray}
\int_{t_1}^{t_2} \int_0^1 \frac{1}{k}\left(\mathcal{E}'' + \frac{1}{r} \mathcal{E}' \right) \left(\mathcal{E}-2\mathcal{B} \right) r dr dt + \int_{t_1}^{t_2} \int_0^1 \frac{1}{k} \left(-\frac{2}{r} \mathcal{E}' + \frac{4}{r} \mathcal{B}' \right)\left(\mathcal{E}-2\mathcal{B}\right) r dr dt \nonumber \\ = \int_{t_1}^{t_2} \frac{1}{k} r \mathcal{E}' \left(\mathcal{E}-2\mathcal{B}\right)  \Big|^{r=1}_{r=0} dt - \int_{t_1}^{t_2} \int_0^1 \frac{1}{k} r \left(\left(\mathcal{E}^\prime\right)^2- 2 \mathcal{B}' \mathcal{E}' \right)dr dt - \int_{t_1}^{t_2} \frac{1}{k} \left(\mathcal{E} - 2 \mathcal{B}\right)^2 \Big|^{r=1}_{r=0} dt\nonumber \\
\end{eqnarray}
The first boundary term vanishes because $\mathcal{E}$ satisfies Neumann conditions at $r=1$ and manifestly vanishes at $r=0$. The second boundary term does not vanish but since $\mathcal{B}$ satisfies Dirichlet conditions at $r=1$ and the conical condition is satisfied at $r=0$ it equals 
\begin{equation}
- \int_{t_1}^{t_2} \frac{1}{k} \left(\mathcal{E} - 2
  \mathcal{B}\right)^2 \Big|^{r=1}_{r=0} dt = - \int_{t_1}^{t_2} \frac{1}{k} \mathcal{E}^2 \left(t,1\right) dt \leq 0 
\end{equation}
Note that without the conical condition imposed, the sign of this term would not be controlled and the uniqueness proof would fail. \\
The other terms appearing on the right hand side in the process of the multiplication and integration will be estimated as follows. For the difference of the exponentials we observe
\begin{eqnarray} \label{expo}
e^{4B+2C-2E} &-& e^{4\tilde{B}+2\tilde{C}-2\tilde{E}}  = \int_0^{1} \frac{\partial}{\partial{\tau}} \left[e^{\left(4B+2C-2E\right) \tau + \left(1-\tau\right)\left(4\tilde{B}+2\tilde{C}-2\tilde{E}\right)}\right] d\tau \nonumber \\ &=&  \left[\int_0^{1} e^{\left(4B+2C-2E\right) \tau + \left(1-\tau\right)\left(4\tilde{B}+2\tilde{C}-2\tilde{E}\right)} d\tau \right] \left(4\mathcal{B} + 2\mathcal{C} - 2\mathcal{E} \right) \nonumber \\ &\leq&  C
 \left(|\mathcal{B}| +  |\mathcal{C}| + |\mathcal{E}|\right)
\end{eqnarray}
for some constant $C$, following from the fact that the solutions
$(B,C,E)$ and $(\tilde{B}, \tilde{C}, \tilde{E})$ are
bounded. Consequently, the remaining terms on the right hand side of
the $\mathcal{B}$ equation when multiplied by $\left(\mathcal{B}r +
\dot{\mathcal{B}}r\right)$ can be estimated (note that the expression
$e^{4B+2C-2E}$ is bounded above and below by some constant)
\begin{equation} \label{bspl}
rem_\mathcal{B} \leq K \int_{t_1}^{t_2} \int_0^1 r \left(\mathcal{B}^2
+ \mathcal{C}^2 + \mathcal{E}^2\right) dr dt + \epsilon
\int_{t_1}^{t_2} \int_0^1 r e^{4B+2C-2E} \left(\dot{\mathcal{B}}^2\right) dr dt \, .
\end{equation}
Here we frequently applied Cauchy's inequality
\begin{equation} \label{Cauchy}
ab \leq \frac{\nu}{2} a^2 + \frac{1}{2\nu} b^2 \, ,
\end{equation}
by means of which the $\epsilon$ in (\ref{bspl}) can be made arbitrarily small on the cost of $K$ becoming larger and larger. By the same token -- since the remaining terms of the $\mathcal{C}$ equation are just given by differences of exponentials of the type considered in (\ref{expo}) -- we can estimate
\begin{equation}
rem_\mathcal{C} \leq K \int_{t_1}^{t_2} \int_0^1 r \left(\mathcal{B}^2 + \mathcal{C}^2 + \mathcal{E}^2\right)
\end{equation}
For the remainder terms of the $\mathcal{E}$ equation (multiplied by $\frac{1}{k}r \left(\mathcal{E} - 2 \mathcal{B} \right)$) we estimate
\begin{eqnarray} \label{delest}
\Bigg|  \frac{1}{k}\left(\mathcal{E} - 2 \mathcal{B}\right)r&& \Bigg(\left[C+\tilde{C}\right]'\mathcal{C}^\prime
+2\left[B+\tilde{B}\right]'\mathcal{B}^\prime \nonumber \\
&& \phantom{XX} - \left[2B+C-E +
  2\tilde{B} + \tilde{C} - \tilde{E}\right]^\prime
\left(2\mathcal{B}^\prime + \mathcal{C}^\prime -
\mathcal{E}^\prime\right)\Bigg) \Bigg| \nonumber \\ && \leq
\delta \cdot r \left({\mathcal{E}^\prime}^2 +{\mathcal{B}^\prime}^2 + {\mathcal{C}^\prime}^2
\right) + K  \cdot r \left(\mathcal{E}^2 + \mathcal{B}^2+\mathcal{C}^2\right) 
\end{eqnarray}
where we have used inequality (\ref{Cauchy}) and the fact that the spatial derivatives of the solutions are bounded. Taking the other terms (again differences of exponentials) into account we arrive at the estimate
\begin{equation}
rem_\mathcal{E} \leq K \int_{t_1}^{t_2} \int_0^1 r \left(\mathcal{B}^2
+ \mathcal{C}^2 + \mathcal{E}^2\right) dr dt + \delta \int_{t_1}^{t_2}
\int_0^1 r \left(\mathcal{B'}^2 + \mathcal{C'}^2 +
\mathcal{E'}^2\right) dr dt
\end{equation}
where $K$ is some constant. Adding up the equations we find the inequality
\begin{eqnarray} \label{ep2}
 \frac{1}{2} \int_0^1 r e^{4B+2C-2E} \left[
  \mathcal{C}^2 + \frac{1}{2k} \mathcal{E}^2 + \frac{1}{2k} \left(\mathcal{E}-4\mathcal{B}\right)^2 + \left(1-\frac{8}{k}\right) \mathcal{B}^2 \right]dr \Big|_{t=t_1}^{t=t_2}  \nonumber \\
 + \frac{1}{2} \int_0^1 
  {\mathcal{B}'}^2 r dr \Big|_{t=t_1}^{t=t_2} + \int_{t_1}^{t_2} \int_0^1 r \left(1-\frac{1}{k} - \epsilon\right) e^{4B+2C-2E} {\dot{\mathcal{B}}}^2 dr dt \nonumber \\
+ \int_{t_1}^{t_2} \int_0^1 r \left(\frac{1}{2k} \left(\mathcal{E}^\prime- 2\mathcal{B}' \right)^2 + \left(\frac{1}{2k}-\delta\right) \left(\mathcal{E}^\prime\right)^2 + \left(1-\frac{2}{k}-\delta\right) {\mathcal{B}'}^2+ \left(1-\delta\right)\mathcal{C'}^2 \right)dr dt \nonumber \\
\leq \hat{K} \int_{t_1}^{t_2} \int_0^1 r \left(\mathcal{B}^2 +
  \mathcal{C}^2 + \mathcal{E}^2\right) dr dt
\end{eqnarray}
We fix $k$ first (say $k=16$), then choose
$\epsilon, \delta$ small enough (say $\epsilon=\delta=\frac{1}{64}$)
to make all terms on the left hand side of expression (\ref{ep2})
manifestly non-negative. This fixes the constant $\hat{K}$ because
(\ref{Cauchy}) has to be applied with a certain weight to produce the
chosen $\epsilon$ and $\delta$ in the estimates (\ref{bspl}) and (\ref{delest}). Since the time-dependent, manifestly non-negative expression
\begin{eqnarray} \label{pose}
 P\left(t\right) &=&  \frac{1}{2} \int_0^1 r e^{4B+2C-2E} \left[
  \mathcal{C}^2 + \frac{1}{2k} \mathcal{E}^2 + \frac{1}{2k} \left(\mathcal{E}-4\mathcal{B}\right)^2 + \left(1-\frac{8}{k}\right) \mathcal{B}^2 \right]dr  \nonumber \\ &+& \frac{1}{2} \int_0^1 {\mathcal{B}'}^2 r dr 
\end{eqnarray}
is zero at $t_1=0$ we can find an interval $[t_1,\tilde{t}]$ with
$\tilde{t} \leq T$ in which expression (\ref{pose})
is monotonically increasing or constant in time. We then have
\begin{eqnarray} \label{fin}
\max_{t \in [t_1,t_2]} P\left(t\right) \leq \hat{K}  \int_{t_1}^{t_2} \int_0^1 r \left(\mathcal{B}^2 +
  \mathcal{C}^2 + \mathcal{E}^2\right) dr dt \leq \tilde{K}
  \int_{t_1}^{t_2} P\left(t\right) dt \leq \tilde{K}
  \left(t_2-t_1\right)\max_{t \in [t_1,t_2]} P\left(t\right) \nonumber
  \\
\end{eqnarray}
for some constant $\tilde{K}$. Inequality (\ref{fin}) is valid for all $t_2 \leq \tilde{t}$. 
Finally, decomposing the time-interval $\left[t_1,\tilde{t}\right]$ into a finite number of
subintervals, each of size $\tilde{K}\Delta t = \frac{1}{2} $, we find that $P\left(t\right)=0$ for each subinterval and since $P\left(t\right)$ is manifestly non-negative, the solution has to be zero on each such subinterval. But if $\tilde{t}$ was not $T$, expression (\ref{pose}) must be strictly monotonically 
decreasing on a following small interval $[\tilde{t},\tilde{t}+\epsilon]$, which is impossible 
since it was just shown that the manifestly non-negative expression
(\ref{pose}) is zero at $t=\tilde{t}$. Hence $\tilde{t}=T$ and
$\mathcal{B}=\mathcal{C}=\mathcal{E}=0$ everywhere on $[0,1] \times
[0,T]$. \\ \\
We note that a similar calculation can be performed for the
equations following from the NUT-ansatz
\begin{equation} \label{nuta}
g = e^{2A\left(t,r\right)}dr^2 + r^2 e^{2B\left(t,r\right)} \left(\sigma_1^2 + \sigma_2^2 \right) +
r^2 e^{2C\left(t,r\right)} \sigma_3^2
\end{equation}
with the gauge
\begin{equation}
\xi = \left[-\left(2B^\prime\left(t,r\right) + C^\prime\left(t,r\right) -A^\prime\left(t,r\right) \right) +
  \left(2B_0^\prime\left(r\right) + C_0^\prime\left(r\right) - A_0^\prime\left(r\right)\right) \right] dr
\end{equation}
where $A_0\left(r\right)$, $B_0\left(r\right)$ and $C_0\left(r\right)$
are the functions chosen to obtain the Ricci-flat Taub-NUT metric in
(\ref{nuta}).


\begin{thebibliography}{1}

\bibitem{Knopf}
  B.~Chow, D.~Knopf, The Ricci-Flow: An Introduction, AMS Surv. a. Monogr., Vol 110 (2004)

\bibitem{deTurck}
D.~M.~DeTurck,
{\it J. Differential Geom.} {\bf 18}  (1983),  no. 1, 157--162

\bibitem{Fornberg}
B.~Fornberg, 
{\it SIAM J. Sci. Comput.} {\bf 16} (1995) no.~5, 1071-1081

\bibitem{Gibbons:1979xm}
  G.~W.~Gibbons and S.~W.~Hawking,
  {\it Commun.\ Math.\ Phys.\  {\bf 66} (1979) 291.}

\bibitem{Gibbons:1978ac}
  G.~W.~Gibbons, S.~W.~Hawking and M.~J.~Perry,
  {\it Nucl.\ Phys.\ B} {\bf 138} (1978) 141.

\bibitem{Hamilton}
R.~S.~Hamilton, 
{\it J. Differential Geom.}  {\bf17}  (1982), no. 2, 255--306.

\bibitem{Hamilton2}
R.~S.~Hamilton, 
{\it Surv. Differential Geom.} (1995), vol 2, 7--136.

\bibitem{Wiseman}
  M.~Headrick and T.~Wiseman,
  Class.\ Quant.\ Grav.\  {\bf 23} (2006) 6683
  [arXiv:hep-th/0606086].

\bibitem{Lady}
  O.~A.~Lady\v{z}enskaja, V.~A.~Solonnikov, N.~N.~Ural'ceva, Linear and
  Quasilinear Equations of Parabolic Type, Transl. Math. Monographs
  {\bf 23}, Providence 1968

\bibitem{Perelman1}
  G.~Perelman,
  arXiv:math.dg/0211159.

\bibitem{Perelman2}
  G.~Perelman,
  arXiv:math.dg/0303109.

\bibitem{Shen}
Y.~Shen, 
 {\it Pacific J. Math.}  {\bf 173}  (1996),  no. 1, 203--221.

\bibitem{Trefethen}
L.~N.~Trefethen, Spectral methods in {MATLAB}, {SIAM},
{Philadelphia},{2000}


\bibitem{Warnick}
  C.~Warnick,
  Class.\ Quant.\ Grav.\  {\bf 23} (2006) 3801
  [arXiv:hep-th/0602127].

\bibitem{Young:1984dn}
  R.~E.~Young,
  Phys.\ Rev.\  D {\bf 28} (1983) 2420.

\end{thebibliography}
\end{document}